# CASCADES OF SUBHARMONIC STATIONARY STATES
# IN STRONGLY NON-LINEAR DRIVEN PLANAR SYSTEMS

Vasyl P. Lukomsky and Ivan S. Gandzha

*Department of Theoretical Physics*, *Institute of Physics*, *National Academy of Sciences,*
*Prospect Nauky* 46, *Kyiv*, *Ukraine*
e-mails: lukom@iop.kiev.ua and gandzha@iop.kiev.ua

Postal address: Dr. Vasyl P. Lukomsky, Institute of Physics, Department of Theoretical Physics, Prospect Nauky 46, 03028, Kyiv, Ukraine

The dynamics of a one-degree of freedom oscillator with arbitrary polynomial non-linearity subjected to an external periodic excitation is studied. The sequences (cascades) of harmonic and subharmonic stationary solutions to the equation of motion are obtained by using the harmonic balance approximation adapted for arbitrary truncation numbers, powers of non-linearity, and orders of subharmonics. A scheme for investigating the stability of the harmonic balance stationary solutions of such a general form is developed on the basis of the Floquet theorem. Besides establishing the stable/unstable nature of a stationary solution, its stability analysis allows obtaining the regions of parameters, where symmetry-breaking and period-doubling bifurcations occur. Thus, for period-doubling cascades, each unstable stationary solution is used as a base solution for finding a subsequent stationary state in a cascade. The procedure is repeated until this stationary state becomes stable provided that a stable solution can finally be achieved. The proposed technique is applied to calculate the sequences of subharmonic stationary states in driven hardening Duffing's oscillator. The existence of stable subharmonic motions found is confirmed by solving the differential equation of motion numerically by means of a time-difference method, with initial conditions being supplied by the harmonic balance approximation.

## 1. INTRODUCTION

Harmonic and subharmonic driven oscillations of the one-degree of freedom non-linear system governed by the following equation of motion are considered

$$\ddot{u} + g\dot{u} + \frac{\partial V}{\partial u} = 2F\cos\omega\, t, \quad V(u) = \sum_{p=1}^{P} \frac{\alpha_p}{p+1} u^{p+1}; \quad u(0) = x_0, \ \dot{u}(0) = y_0; \qquad (1.1)$$

where a dot designates a derivative with respect to time; $\alpha_p$, $x_0$, $y_0$ are arbitrary real constants; $g$, $F$, $\omega$ are positive real constants; $1 \le P < \infty$.

In many physical systems, especially in the case of strong external excitation, non-linear oscillations are described by the equations of the form (1.1) with polynomial or exponential non-linearity [1–3]. In this way, the potential of the interionic interaction in solid bodies is approximately described by the Lenard-Jones model $V(u) = -2u^{-6} + u^{-12}$ [4]. In the nearest neighbors model for a chain of atoms, Toda's potential $V(u) = e^u - u$ [5] describes the exponential growth in the repulsion of atoms as they approach, and the constant force of the attraction as they recede. In the case $V(u) = \cos u - 1$, equation (1.1) describes the oscillations of the mathematical pendulum. When studying non-linear gravity waves on a fluid surface, equations with exponential non-linearity also arise after the conformal mapping of the domain filled by fluid to the unity circle [6]. Multiple-harmonic generation and sum-frequency generation in a non-linear crystal subjected to laser radiation are due to the polynomial non-linearity of polarization in electric field strength [7].

In this paper, steady periodic solutions of equation (1.1) are studied. Since the system is driven by the external periodic excitation stationary solutions do not depend on the initial conditions $\{x_0,\ y_0\}$. Initial conditions only define what stationary solution from all the possible ones is chosen by the oscillator after finishing transient motion. Note, however, that not all initial conditions lead to periodic steady motion. In general, there are three kinds of spatially bounded steady motion in non-linear systems: periodic, quasi-periodic, and chaotic [8]. Quasi-periodic motion arises in systems, where two or more governing frequencies are present, with the ratio being an irrational number. These could be multi-frequency excited systems or single-frequency excited systems admitting limit cycles. Therefore, quasi-periodic motion is not possible in single-frequency excited systems without limit cycles such as system (1.1). In this case, initial conditions can lead only to periodic or chaotic motion. Chaotic motion is bounded stochastic non-periodic motion of deterministic systems that is characterized by a number of distinctive features [9].

There are two main approaches for finding stationary solutions to the equation of form (1.1) without assuming non-linearity to be small. The first one (that is applicable only to stable solutions) is to solve it numerically for a variety of initial conditions using one of the time-difference methods [10] (the Runge-Kutta, Adams, etc. methods) waiting every time until motion settles down onto a steady oscillation. Besides inconvenience of searching between different initial conditions, the main technical problem here is that numerical errors accumulate with time, especially in the vicinity of the boundaries of chaotic motion. This highly complicates the chance to find all the possible stationary solutions for particular values of parameters.

If transient processes are not of interest then the second and a more reliable way is to find stationary solutions (both stable and unstable) at first and after that to calculate what initial conditions are needed to realize each of stable stationary periodic states of motion. For this purpose, the harmonic balance method [3], which is widely used to study strongly non-linear systems [11−16], seems to be most effective (it is also often called by Gallerkin's method [17]). The main advantage of the harmonic balance method over time-difference methods is that it allows obtaining both stable and unstable stationary solutions [12]. Note that the shooting method with continuation technique also allows computing unstable solutions [18, 19]. Contrasting the harmonic balance method to shooting method, however, is beyond the scope of the present work. Some remarks on this topic can be found in the two papers cited above and also in [11].

Let $\bar{u}^{(0)}(t)$ be some stationary periodic solution of equation (1.1) describing harmonic oscillations, whose period is equal to the period of the external excitation. In the harmonic balance method, this solution is presented by the following truncated Fourier series

$$\bar{u}^{(0)}(t) = \sum_{n=-N}^{N} \bar{u}_n^{(0)} e^{in\omega t}, \quad \bar{u}_{-n}^{(0)} \equiv \bar{u}_n^{(0)^*}, \tag{1.2}$$

where $N$ is the number of the highest harmonic taken into account in the approximate solution; (*) is the complex conjugate; the dash means that a solution is stationary, its harmonics not depending on time in this case.

After substitution of (1.2) into (1.1), the complex harmonics $\bar{u}_n^{(0)}$ are defined by the following system of non-linear algebraic equations

$$f_n(\omega)\,\overline{u}_n^{(0)} + \sum_{p=1}^{P} \alpha_p (\overline{u}^{(0)p})_n = F\delta_{|n|,1}, \quad |n| = \overline{0, N}, \tag{1.3}$$

where $f_n(\omega) = -n^2\omega^2 + ign\omega$, $\delta_{|n|,1}$ is the Kronecker delta. The harmonics $(\overline{u}^{(0)p})_n$ are the polynomials of power $p$ in the harmonics $\overline{u}_n^{(0)}$. The corresponding recurrence expressions are presented in Section 2.1 or [20], and their explicit form is given in Appendix A. Using these recurrence formulae and complex exponents in Fourier expansions is the key novel step in the formulation of the harmonic balance method developed in this paper.

System (1.3) is equivalent to differential equation (1.1) at $N \to \infty$, in the class of the periodic solutions with main harmonic of frequency $\omega$. In a particular range of parameters, however, the equation of motion (1.1) admits more complicated solutions with periods $m$ times greater than the period of the external excitation, $m$ being a positive integer. Such solutions are called subharmonic of order $m$ and often arise as a result of instability of harmonic solutions [1]. The harmonic balance procedure for obtaining such solutions with arbitrary number of modes is given in Sections 2.1 and 2.2. There, each subharmonic solution with prime multiplicity (a first level solution) is looked for as a deviation from a harmonic solution (a zero level solution), which is called a base solution in this case. Then, the subharmonic solutions found are used as the base ones for obtaining subharmonic solutions with higher multiplicities being the product of two prime numbers (second level solutions), and so on. In such a way, the sequences (cascades) of stationary states corresponding to subharmonic oscillations with periods of increasing multiplicities are constructed. When the deviation of some stationary solution from a base one is infinitesimal at the boundary of the region of parameters, where this stationary solution is excited (new harmonics excited are zero at bifurcation points), the excitation is called *soft* (a supercritical bifurcation). Otherwise, the excitation is called *hard* (a subcritical bifurcation).

The stable/unstable nature of the harmonic balance solutions is usually decided by using the Floquet theory [1]. However, its application is usually restricted to a few leading harmonics due to the awkwardness of the procedure [14, 15]. In Section 2.3, a generalized notation for the scheme of stability analysis based on the Floquet theorem [21] and suitable for arbitrary truncation numbers is proposed using complex exponents instead of sines and cosines in Fourier series. The following consequence of the Floquet theorem for second-order differential equations is emphasized. If a stationary solution is unstable then there are three possibilities: (i) a solution is *metastable* (e.g., the middle branch in hysteresis) and could not physically be realized (such solutions are often associated with resonant jumps due to a saddle-node bifurcation); (ii) new harmonics are excited without changing the period of a solution due to a symmetry-breaking bifurcation; (iii) the period of a solution is doubled due to a period-doubling bifurcation. We also point out that it is possible to investigate stability with respect to period-doubling or symmetry-breaking without including the harmonics to be excited into the variation of an unstable solution. Thus, each unstable stationary solution in a period-doubling cascade is used for finding the boundaries of a subsequent stationary state to which an oscillator proceeds after a period-doubling bifurcation. Using this procedure successively to each unstable solution allows describing the whole cascade of subharmonic solutions with doubling periods, until a stable solution is achieved. If a stable state cannot be achieved at particular values of parameters then a system of type (1.1) should exhibit a non-periodic chaotic response.

It follows from the Floquet theorem that a period-doubling bifurcation is the only way for a system of type (1.1) to change the period of oscillations without any jump in amplitude (it is not possible to trace a period-multiplying bifurcation of odd order by the infinitesimal Floquet theory). Therefore, the following conjecture is made: excitation of odd-order subharmonics is to all appearance hard, whereas excitation of even-order subharmonics is to all appearance soft. In Sections 3 and 4, this conjecture is verified on the example of driven *hardening* Duffing's oscillator, where both softly (Section 3) and hardly (Section 4) excited stable and unstable subharmonic solutions were calculated and several period-doubling cascades were obtained in the region between the second and the third superharmonic resonances. Note that on the contrary to the softening Duffing equation (which is well studied, e.g., in [12, 16, 18]), the subharmonic motion in the *hardening* Duffing equation was not almost studied at all. The difficulty here is that subharmonic oscillations are strongly non-linear since they appear in the region of superharmonic resonances and demand much more computational efforts than in the case of the softening Duffing oscillator.

By using initial conditions supplied by the harmonic balance approximation stable subharmonic solutions of the hardening Duffing equation were also verified numerically by means of the time-difference Runge-Kutta method. Infinite period-doubling cascades were traced up to chaotic attractors. Chaotic motion in the hardening Duffing oscillator is shown to coexist with the stable third superharmonic resonant solution and, therefore, is not associated with the loss of its stability (as was assumed in the papers [24, 15]) but with the existence of infinite period-doubling cascades. Abrupt transition to/from chaotic motion via transient chaos without a period-doubling cascade was found to occur at the point, where the amplitude-frequency characteristics of two stable and unstable hardly excited subharmonic solutions of the same order intersect and change their stability due to a transcritical bifurcation. Concluding comments are given in Section 5.

# 2. MODELING STEADY SUBHARMONIC OSCILLATIONS

## 2.1. THE HARMONIC BALANCE APPROXIMATION FOR SUNBHARMONIC OSCILLATIONS

Let the stationary solution $\overline{u}^{(0)}(t)$ presented by expansion (1.2) be known. The function $\overline{u}^{(0)}(t)$ is further called a *zero level solution*. It satisfies the differential equation

$$\ddot{\overline{u}}^{(0)} + g\dot{\overline{u}}^{(0)} + \sum_{p=1}^{P} \alpha_p \overline{u}^{(0)^p} = 2F\cos\omega t \,. \tag{2.1}$$

Let $u(t)$ be some other, not obligatory stationary, solution of equation (1.1). Denote the deviation of this solution from the zero level stationary solution as $\upsilon(t)$:

$$u(t) = \overline{u}^{(0)}(t) + \upsilon(t) \,. \tag{2.2}$$

Substitution of (2.2) into initial equation (1.1) with taking into account (2.1) leads to the following homogeneous non-linear differential equation for the deviation $\upsilon(t)$

$$\ddot{\upsilon} + g\dot{\upsilon} + \sum_{p=1}^{P} \left(\alpha_p + \beta_p^{(0)}(t)\right)\upsilon^p = 0, \tag{2.3}$$

where
$$\beta_p^{(0)}(t) = \sum_{p_1=1}^{P-p} C_{p+p_1}^p \alpha_{p+p_1} \overline{u}^{(0)^{p_1}}(t) \,, \quad \beta_P^{(0)}(t) \equiv 0 \,; \tag{2.4}$$

$C_{p+p_1}^p$ being the binomial coefficients. The linearized version of this equation is called a variational equation [1]. In this case, its solutions describe the dynamics of a system under small deviations from the zero level stationary solution.

In a particular range of parameters, other stationary solutions (stable or unstable) corresponding to subharmonic oscillations with period divisible by the period $T$ of the external excitation may exist along with the zero level stationary solution $\overline{u}^{(0)}(t)$ (given by expansion (1.2)), especially when it is unstable. The deviation of such a subharmonic solution from the zero level stationary solution satisfies equation (2.3). Let $\upsilon^{(1)}(t; m)$ be the solution of this equation with period $mT$, $m$ being a prime number ($m = 2, 3, 5, 7, \ldots$). Then the function

$$u^{(1)}(t; m) = \overline{u}^{(0)}(t) + v^{(1)}(t; m) \qquad (2.5)$$

is the solution of initial equation (1.1). Such a solution is further called a *first level solution*.

Limiting ourselves by the frequency bound $N\omega$, as in expansion (1.2), the deviation $v^{(1)}(t; m)$ is looked for as the following truncated Fourier series

$$v^{(1)}(t; m) = \sum_{n=-mN}^{mN} v^{(1)}_{\frac{n}{m}}(t) \, e^{i\frac{n}{m}\omega t}, \quad v^{(1)}_{-\frac{n}{m}}(t) \equiv v^{(1)*}_{\frac{n}{m}}(t). \qquad (2.6)$$

The integer powers $v^p$ of the function $v$ are also expressed in the form of the truncated Fourier series

$$v^p(t; m) = \sum_{n=-mpN}^{mpN} (v^p)_{\frac{n}{m}} \, e^{i\frac{n}{m}\omega t}, \quad p = \overline{1, P}. \qquad (2.7)$$

Using complex exponents in expansions (2.6) and (2.7) makes it possible to express the harmonics $(v^p)_{\frac{n}{m}}$ recurrently in terms of the harmonics $v_{\frac{n}{m}}$ (the term "harmonics" hereafter includes both the harmonics with integer indices and the subharmonics with fractional indices)

$$(v^p)_{\frac{n}{m}} = \sum_{n_1 = \max(-(p-1)mN, \, n-mN)}^{\min((p-1)mN, \, n+mN)} (v^{p-1})_{\frac{n_1}{m}} v_{\frac{n-n_1}{m}}, \quad |n| = \overline{0, mpN}; \quad p = \overline{2, P}. \qquad (2.8)$$

This formula follows from the expression for the harmonics of the product of two periodic functions given in Appendix A. The explicit expression for $(v^p)_{\frac{n}{m}}$ is also presented there.

Recurrence formulae (2.8) were not used before to the knowledge of the authors and represent the key point in the proposed formulation of the harmonic balance method. One can at once notice the advantage over the awkward algebra presented in a variety of publications (see, e.g., [13, 25]), where Fourier expansions in terms of sines and cosines are used.

Obtaining recurrence formulae turned out to be possible in that case too [26], but advantage and simplicity of (2.8) is apparent. Another way to calculate the harmonics $(v^p)_{\frac{n}{m}}$ is to use the Fast Fourier Transform (FFT), as it is performed by a number of authors in [11, 12]. Comparing the numerical efficiency of FFT and recurrence formulae (2.8) is left for future investigation, however, (2.8) is obviously much more useful for analytic implementation in conjunction with analytic computer algebra.

Substitution of (2.6) and (2.7) into equation (2.3) leads to the following system of differential equations for the harmonics $v_{\frac{n}{m}}^{(l_1)}(t)$ $(l_1 = 1)$

$$\ddot{v}_{\frac{n}{m}}^{(l_1)} + (2i\frac{n}{m}\omega + g)\dot{v}_{\frac{n}{m}}^{(l_1)} + \sum_{n_1 = N_n^-(1;\,m)}^{N_n^+(1;\,m)} \left( (f_{\frac{n_1}{m}} + \alpha_1)\,\delta_{n,\,n_1} + \beta_{1,\,\frac{n-n_1}{m}}^{(l_0)} \right) v_{\frac{n_1}{m}}^{(l_1)} +$$
$$+ \sum_{p=2}^{P} \sum_{n_1 = N_n^-(p;\,m)}^{N_n^+(p;\,m)} \left( \alpha_p \delta_{n,\,n_1} + \beta_{p,\,\frac{n-n_1}{m}}^{(l_0)} \right) (v^{(l_1)\,p})_{\frac{n_1}{m}} = 0, \quad |n| = \overline{0,\,mN}; \tag{2.9}$$

$$\beta_{p,\,\frac{n}{m}}^{(l_0)} = \sum_{p_1=1}^{P-p} \alpha_{p+p_1} C_{p+p_1}^{p} (\overline{u}^{(l_0)\,p_1})_{\frac{n}{m}}, \quad p = \overline{1,\,P-1}; \quad \beta_{P,\,\frac{n}{m}}^{(l_0)} \equiv 0; \tag{2.10}$$

where the designations $l_0 = 0$ and $l_1 = 1$ are used for convenience;

$$N_n^-(p;\,m) = \max\left(-mpN,\,n - m(P-p)N\right),\ N_n^+(p;\,m) = \min\left(mpN,\,n + m(P-p)N\right).$$

Since the zero level periodic solution contains the harmonics with only the integer numbers, $\beta_{p,\,\frac{n}{m}}^{(0)} = 0$ when $n$ is not divisible by $m$. Equations (2.9) with $n$ divisible by $m$ describe a reaction of the zero level stationary solution to excitation of the subharmonics of order $m$.

## 2.2. SUBHARMONIC SOLUTIONS OF HIGHER ORDERS

Use the described above principle of constructing the zero and first level stationary solutions of system (1.1) for finding higher order stationary solutions of subsequent levels. The model for multi-level cascades of subharmonic solutions is shown in Figure 1. The point $A_1$ in the center of the circles corresponds to the zero level solution $\bar{u}^{(0)}(t)$. Its spectrum contains only the integer harmonics with main frequency $\omega$. The first level solutions $\bar{u}^{(1)}(t;m)$ with period $mT$, $m$ being a prime number ($m = 2, 3, 5, 7, \ldots$), are located on the first circle and are presented by the points $A_m$. Further, each of the first level states $A_m$ found can be considered similarly to the zero level state $A_1$ for obtaining second level states with multiplicity being the product of two prime numbers. Such stationary states are located on the second circle in Figure 1. Similarly, stationary states with multiplicity being the product of three prime numbers are located on the third circle, and so on. In this way, all the possible stationary solutions to the equation of motion can be traced. The procedure is not unique. For example, the second level state $A_6$ can be formed by two ways: $A_1 \to A_2 \to A_6$ or $A_1 \to A_3 \to A_6$.

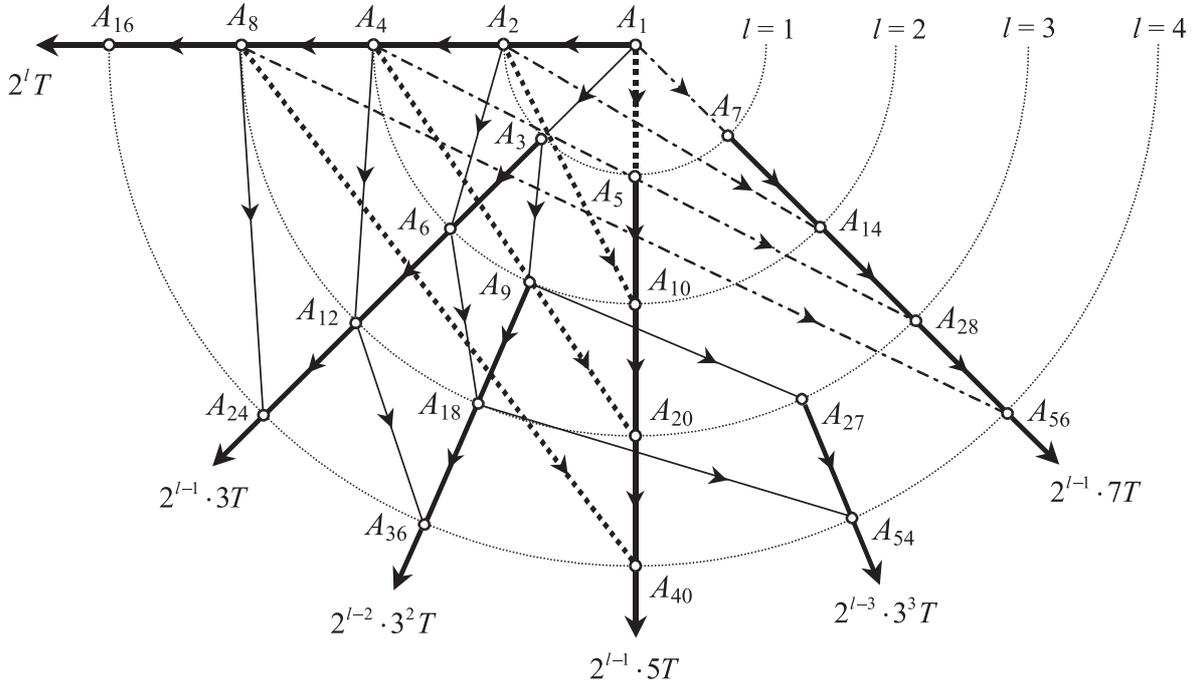

**Figure 1.** The diagram for obtaining the cascades of subharmonic states. The states containing the subharmonics of order $m$ are presented by the points $A_m$. ▬▬▬ , Period-doubling; ——— , period-tripling; ▪▪▪▪▪▪ , fivefold period-multiplying; –·–·– , sevenfold period-multiplying.

Thus, let the stationary solution $\bar{u}^{(l_0)}(t; m_0)$ of level $l_0$ with period $m_0 T$ be known, $m_0$ being the product of $l_0$ prime numbers. Then the stationary solution $\bar{u}^{(l_1)}(t; m)$ of subsequent level $l_1 = l_0 + 1$ with period $mT$ ($m = m_0 r$, $r$ being a prime number) is looked for as a deviation from the solution $\bar{u}^{(l_0)}(t; m_0)$:

$$\bar{u}^{(l_1)}(t; m) = \bar{u}^{(l_0)}(t; m_0) + \bar{\upsilon}^{(l_1)}(t; m). \tag{2.11}$$

Then the harmonics of the deviation $\bar{\upsilon}^{(l_1)}(t; m)$ are found from the following system of non-linear algebraic equations being the stationary version of system (2.9):

$$
\begin{aligned}
\sum_{n_1 = N_n^-(1; m)}^{N_n^+(1; m)} & \left( (f_{\frac{n_1}{m}} + \alpha_1) \, \delta_{n, n_1} + \beta^{(l_0)}_{1, \frac{n-n_1}{m}} \right) \bar{\upsilon}^{(l_1)}_{\frac{n_1}{m}} + \\
& + \sum_{p=2}^{P} \sum_{n_1 = N_n^-(p; m)}^{N_n^+(p; m)} \left( \alpha_p \delta_{n, n_1} + \beta^{(l_0)}_{p, \frac{n-n_1}{m}} \right) (\bar{\upsilon}^{(l_1) p})_{\frac{n_1}{m}} = 0, \quad |n| = \overline{0, mN};
\end{aligned}
\tag{2.12}
$$

The coefficients $\beta^{(l_0)}_{p, \frac{n}{m}}$ are different from zero only if $n$ is divisible by $r$ ($m = m_0 r$) and are defined by formula (2.10).

## 2.3. Stability analysis

To analyse the stability of a stationary periodic solution to equation (1.1) the linearized version of equation (2.3) (the variational equation) for the infinitesimal variation of this solution is usually used [1] (any stationary solution can be substituted into (2.3) instead of a zero level solution). Further analysis is based on the Floquet theorem for linear differential equations with periodic coefficients [21] resulting in the variational equation to have the following partial solution:

$$\upsilon(t) = \exp(i\Omega t) \, \varphi(t) \equiv \exp((-g/2 + \mu) \, t) \, \varphi(t), \; \varphi(t + T) = \varphi(t); \tag{2.13}$$

$T$ being the period of a stationary solution, whose stability is investigated. It is unstable if $\text{Im}\,\Omega < 0$ ($\text{Re}(-g/2 + \mu) > 0$). The characteristic frequency $\Omega$ and the characteristic exponent $\mu$ are in general complex numbers. However, it can be shown (see, e.g., [1]) that the function $\cosh \mu T$ should be real. This leads to three following cases:

(i) $\mu$ is imaginary ($\operatorname{Im}\Omega = g/2$), a stationary solution is stable ($g > 0$);

(ii) $\mu$ is real ($\operatorname{Re}\Omega = 0$, $\operatorname{Im}\Omega = g/2 - \mu$), a solution is unstable (such as the middle branch in hysteresis, that is, a metastable state) if $\operatorname{Im}\Omega < 0$;

(iii) $\mu$ is complex, but $\operatorname{Im}\mu = n\pi/T$ ($\operatorname{Im}\Omega \neq 0$ and $\operatorname{Re}\Omega = n\pi/T$), $n$ being an integer. This case is crucial indicating a period-doubling instability when $\operatorname{Im}\Omega < 0$ and $n$ is odd. Instability with even $n$ is a symmetry-breaking one (when the even harmonics are excited in an unstable stationary solution with only the odd harmonics). Note that the characteristic multiplier $q = \exp(i \operatorname{Im}\mu T)$ is often used (see, e.g., [12, 13]) instead of the characteristic exponent $\mu$. It is seen that $q = -1$ for a period-doubling instability and $q = 1$ for a symmetry-breaking one.

No other possibility except the cases (i), (ii), (iii) is admitted. Therefore, linear stability (Floquet) analysis is not capable of tracing period-multiplying bifurcations of odd order. Bearing in mind that the Floquet theory deals with only the infinitesimal variations one comes to the following conjecture/hypothesis valid for the systems of type (1.1): excitation of odd-order subharmonics is to all appearance hard (subcritical), whereas excitation of even-order subharmonics is to all appearance soft (supercritical). Except of this conclusion, the scheme for stability analysis proposed hereafter does not conceptually bring something new to the Floquet theory. However, several technical innovations introduced make stability analysis based on the Floquet theory more systematic and general. Using complex exponents instead of sines and cosines in Fourier expansions makes it possible to formulate the systematic procedure for arbitrary number of modes and powers of non-linearity on the contrary to many studies [1, 14–16], where only several first approximations can be considered with extremely awkward algebra.

Complex $\mu$ are often not considered [1, 13–16]. Instead of this, the following action is accomplished:

(iii-a) the function $\varphi(t)$ in (2.13) is additionally assumed to have the period $2T$ since the function $\exp(i\,\mathrm{Im}\,\mu\,t) = \exp(in\pi t/T)$ is periodic with periods $T$ or $2T$ and can be included into the function $\varphi(t)$. By making so the case (iii) is reduced to the case with real $\mu$, however, harmonics not presented in a stationary solution should be included into the variation $\upsilon(t)$.

Thus, to investigate the stability of any stationary solution, say, $\overline{u}^{(l_0)}(t; m_0)$ it is sufficient to include into the variation $\upsilon^{(l_1)}(t)$ $(l_1 = l_0 + 1)$ of this solution the subharmonics $\upsilon^{(l_1)}_{\frac{n}{m}}(t)$ of order $m = 2m_0$. In this case, the Floquet theorem (2.13) attains the following form:

$$\upsilon^{(l_1)}_{\frac{n}{m}}(t) = \upsilon^{(l_1)}_{\frac{n}{m}} \exp(i\Omega\,t)\,, \quad |n| = \overline{0, mN}\,; \tag{2.14}$$

$\upsilon^{(l_1)}_{\frac{n}{m}}$ being the harmonics of the function $\varphi(t)$ in (2.13). To perform constructing stationary solutions and analysing their stability in the framework of a single systematic scheme it is more convenient to linearize the system of differential equations (2.9) instead of variational equation (2.3) (these two ways are equivalent). Taking into account (2.14) this leads to the following system of $2mN + 1$ homogeneous linear algebraic equations:

$$\sum_{n_1 = N_n^-(1; m)}^{N_n^+(1; m)} \left( \left( -(\frac{n}{m}\omega + \Omega)^2 + ig(\frac{n}{m}\omega + \Omega) + \alpha_1 \right) \delta_{n, n_1} + \beta^{(l_0)}_{1, \frac{n - n_1}{m}} \right) \upsilon^{(l_1)}_{\frac{n}{m}} = 0, \quad |n| = \overline{0, mN}\,; \tag{2.15}$$

where $m = 2m_0$ and $l_1 = l_0 + 1$. The characteristic frequencies are found by equating the determinant of this system to zero.

Since $\beta^{(l_0)}_{1, \frac{n}{2m_0}} \neq 0$ only when $n$ is even, linear system (2.15) comes apart into two independent subsystems. The first subsystem consists of $2m_0N+1$ equations for the harmonics $\upsilon^{(l_1)}_{\frac{n}{m_0}}$ ($|n| = \overline{0, m_0 N}$) and is obtained from (2.15) by changing $m$ to $m_0$. Only the variations of the harmonics presented in the stationary solution $\bar{u}^{(l_0)}(t; m_0)$ are considered in this case. The stationary solution is unstable if the imaginary part of at least one of the characteristic frequencies $\Omega_l$ ($l = \overline{1, 2m_0N+1}$) turns out to be negative according to the clauses (ii) and (iii). Complex $\Omega_l$ should be considered for tracing period-doubling and symmetry-breaking instabilities, but this allows investigating all the possible instabilities including the saddle-node ones using the same set of equations. Note that a symmetry-breaking instability is possible when a stationary solution includes only the odd harmonics ($\beta^{(l_0)}_{1, \frac{n}{m_0}} \neq 0$ only when $n$ is odd) and, therefore, the first subsystem consists of the equations with only odd $n$. In this case, any period-doubling instability should be preceded by a symmetry-breaking one.

The second subsystem consists of the rest of equations (2.15) not included into the first subsystem. The second subsystem corresponds to the case (iii-a) when the harmonics to be excited are included into the variation, that is, the harmonics $\upsilon^{(l_1)}_{\frac{n}{2m_0}}$ with odd $n$ for a period-doubling instability or the harmonics $\upsilon^{(l_1)}_{\frac{n}{m_0}}$ with even $n$ for a symmetry-breaking instability. Since the frequency of these harmonics is specified in the second subsystem *a priori*, bifurcation points in this case are established more precisely than when using the first subsystem, where this frequency is determined using the clause (iii) only approximately (because the finite number of harmonics is taken into account in (2.14)).

Thus, it is useful to unite the two approaches (iii) and (iii-a) to treat characteristic exponents into a single scheme. Complex characteristic frequencies are considered at first to find period-doubling and symmetry-breaking regions using only the harmonics presented in a stationary solution (system (2.15) with $m = m_0$). After that, the harmonics to be excited are included into the deviation to calculate bifurcation points more precisely (system (2.15) with $m = 2m_0$ and odd $n$ for a period-doubling instability or $m = m_0$ and even $n$ for a symmetry-breaking instability). Finally, to carry out numerical calculations it is convenient to reformulate linear system (2.15) as an eigenvalue problem (see Appendix B).

For period-doubling cascades, the Floquet theory results in that states of higher multiplicity can be achieved only through the states of lower multiplicity. For example, to achieve the fourfold state $A_{16}$ on the ray of period doubling a system must subsequently pass through the states $A_2 \rightarrow A_4 \rightarrow A_8 \rightarrow A_{16}$ (see Figure 1). This rule was at first established in [27] when studying self-mapping of a straight line. Moreover, any period-doubling cascade is governed by a universal law revealed by Feigenbaum [28]. Thus, the stability analysis based on the Floquet theory makes studying period-doubling cascades more systematic. In this case, any unstable solution can be used for finding the regions of parameters, where a subsequent stationary solution with doubled period is excited.

### 2.4. AMPLITUDE DEGENERATION OF SUBHARMONIC SOLUTIONS

Equation of motion (1.1) admits the following transformation of periodicity:

$$u \rightarrow u, \quad t \rightarrow t + kT, \ k \in \mathbb{Z}; \tag{2.16}$$

$T$ being the period of the external excitation. Let $u(t; m)$ be some subharmonic solution with period $mT$ ($m = 1$ corresponds to a harmonic solution). Then due to periodicity (2.16) the equation of motion admits the whole family of $m$ subharmonic solutions of the same order:

$$u(t; m \,|\, k) = u(t + kT; m), \quad k = \overline{0, m-1} \,. \tag{2.17}$$

The same harmonics of all the solutions in a given family are equal in amplitude but different

in phase:

$$u_{\frac{n}{m}}(t\mid k) = u_{\frac{n}{m}}(t\mid 0)\, e^{i2\pi k\frac{n}{m}}, \quad k = \overline{0, m-1}. \tag{2.18}$$

Thus, the subharmonic solutions of order $m$ are all $m$-fold degenerated in amplitude independently of the way of their excitation.

Periodicity rule (2.16) produces an excellent method for investigating driven motion by means of the Poincaré maps (see, e.g., [9]). The location of a solution in the phase plane is recorded every time as the time interval of length $T$ is elapsed starting from $t = 0$. Then the Poincaré map of a stationary subharmonic solution of order $m$ consists of $m$ points. Each of these points is the initial condition to realize one of the solutions from family (2.17).

## 3. THE SOFT EXCITATION OF EVEN-ORDER SUBHARMONIC OSCILLATIONS IN DRIVEN HARDENING DUFFING'S EQUATION

### 3.1. THE SYMMETRY OF SOLUTIONS TO THE EQUATION OF MOTION WITH EVEN POTENTIAL

Hereafter we investigate how the oscillations described by driven hardening Duffing's equation ($P = 3$; $\alpha_2 = 0$)

$$\ddot{u} + 0.2\dot{u} + u + u^3 = 2F\cos\omega t \tag{3.1}$$

depend on the external frequency $\omega$. The equation of motion (3.1) (and, in general, any equation (1.1) with even potential $V(u)$ ($\alpha_{2p} = 0$)) admits the following symmetry transformation

$$u \to -u, \quad t \to t + \frac{T}{2} + kT, \quad k \in \mathbb{Z}; \tag{3.2}$$

where $T = 2\pi/\omega$. Therefore, for any subharmonic solution $u(t; m)$ in this case, there are in general two families of solutions: (2.17) and

$$\hat{u}(t; m \mid k) = -u(t + T/2 + kT; m), \ k = \overline{0, m-1}. \tag{3.3}$$

However, $u(t; m \mid k)$ and $\hat{u}(t; m \mid k)$ represent a single family if the solution $u(t; m)$ is symmetric, that is, $u(t; m) = -u(t + mT/2; m)$. For that, $m$ should be odd and the solution should include only the odd harmonics. Otherwise, the solution $u(t; m)$ is asymmetric and (3.2) leads to a new family of solutions $\hat{u}(t; m \mid k)$, shifted in phase with respect to the family $u(t; m \mid k)$.

## 3.2. PERIOD-DOUBLING CASCADES AND CHAOTIC MOTION

Let $\overline{u}^{(l_0)}(t; m_0)$ be some stationary solution of level $l_0$ containing the subharmonics of order $m_0$. In a particular range of parameters, this solution may become unstable with respect to excitation of higher subharmonics of order $2m_0$ as a result of a period-doubling bifurcation. In this case, one of the characteristic frequencies of linear system (2.15) at $m = m_0$ should be equal to $\Omega \approx \pm \omega/2m_0 - i\varepsilon, \ \varepsilon > 0$ (see the case (iii) in Section 2.3). This fact produces the excellent method for finding the regions of parameters, where a period-doubling bifurcation of the solution $\overline{u}^{(l_0)}(t; m_0)$ occurs. If the solution $\overline{u}^{(l_0)}(t; m_0)$ is unstable with respect to doubling the period then the subsequent stationary solution $\overline{u}^{(l_1)}(t; 2m_0)$ $(l_1 = l_0 + 1)$ has the following form according to (2.11):

$$\overline{u}^{(l_1)}(t; 2m_0) = \overline{u}^{(l_0)}(t; m_0) + \overline{v}^{(l_1)}(t; 2m_0) = \sum_{n=-m_0 N}^{m_0 N} \left( \overline{u}^{(l_0)}_{\frac{n}{m_0}} + \overline{v}^{(l_1)}_{\frac{n}{m_0}} \right) e^{i\frac{n}{m_0}\omega t} + \\ + \sum_{n=-m_0 N}^{m_0 N-1} \overline{v}^{(l_1)}_{\frac{n}{m_0}+\frac{1}{2m_0}} e^{i(\frac{n}{m_0}+\frac{1}{2m_0})\omega t} = \sum_{n=-2m_0 N}^{2m_0 N} \overline{u}^{(l_1)}_{\frac{n}{2m_0}} e^{i\frac{n}{2m_0}\omega t}. \tag{3.4}$$

The harmonics $\overline{v}^{(l_1)}_{\frac{n}{2m_0}}$ ($\mid n \mid = \overline{0, 2m_0 N}$) are found from the system of equations (2.12).

To investigate period-doubling bifurcations in equation (3.1) the zero level harmonic solutions should be found at first. These solutions are well known, the case $F = 10$ having been studied in detail in [14] and $F = 25$ — in [29]. Since the non-linearity is odd, the spectrum of the simplest harmonic stationary solution contains only the odd harmonics almost all over the range of parameter $\omega$ due to symmetry (3.2). This symmetric solution becomes unstable due to a symmetry-breaking bifurcation in a narrow band of external frequencies between the regions of the third superharmonic resonance (when the amplitude of the third harmonic exceeds the amplitude of the first one) and the primary resonance [14, 29]. As a result, the even harmonics are excited (the solution becomes asymmetric) and the second superharmonic resonance is formed [14]. Application of the scheme for stability analysis proposed gives an exact estimation of the region, where the even harmonics are excited. Moreover, the corresponding characteristic frequencies are $\Omega \approx \pm\omega - i\varepsilon, \ \varepsilon > 0$, indicating that the Fourier expansion of a subsequent stationary solution should include the harmonics with frequencies $(2n-1)\omega + \operatorname{Re}\Omega \approx 2n\omega$, that is, the even harmonics.

The amplitude-frequency dependences of the zero level solution $\bar{u}^{(0)}(t)$ containing both the odd and even harmonics are shown in Figure 2 at $F = 10$ and $N = 15$, unstable regions being hatched. The regions 1-2, 5-6 are unstable with characteristic frequency $\Omega = 0 - i\varepsilon, \ \varepsilon > 0$. Such regions of solutions are metastable since they cannot experimentally be realised due to the hysteresis phenomenon [3]. The region 4-3 ($\omega \in [1.198, 1.565]$) is unstable with characteristic frequency $\Omega \approx \pm\omega/2 - i\varepsilon, \ \varepsilon > 0$, indicating the instability with respect to excitation of the second-order subharmonics.

These subharmonics are taken into account in the first level stationary solution $\bar{u}^{(1)}(t; 2)$. The amplitude-frequency dependences of its subharmonics 5/2 and 3/2 are shown in Figure 3. One can see that the region, where the subharmonics are nonzero, coincides with the region of instability shown in Figure 2 (a slight difference is due to the approximation $N = 5$ used instead of $N = 15$, as for the zero level solution). The subharmonic $5/2$ has the greatest amplitude among all the other ones. This indicates that the subharmonic resonance of this order is present. Further analysis shows the obtained first level solution to be stable at $F = 10$.

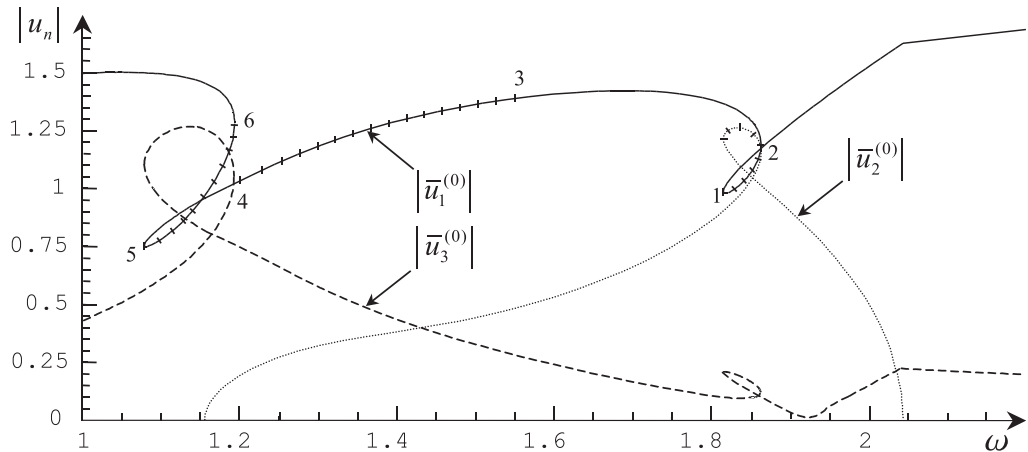

**Figure 2.** The third and the second superharmonic resonances in driven hardening Duffing's oscillator $\ddot{u} + 0.2\dot{u} + u + u^3 = 20\cos\omega t$. The harmonic balance approximation with $N = 15$ was used, both the odd and even harmonics having been taken into account. The unstable regions are hatched.

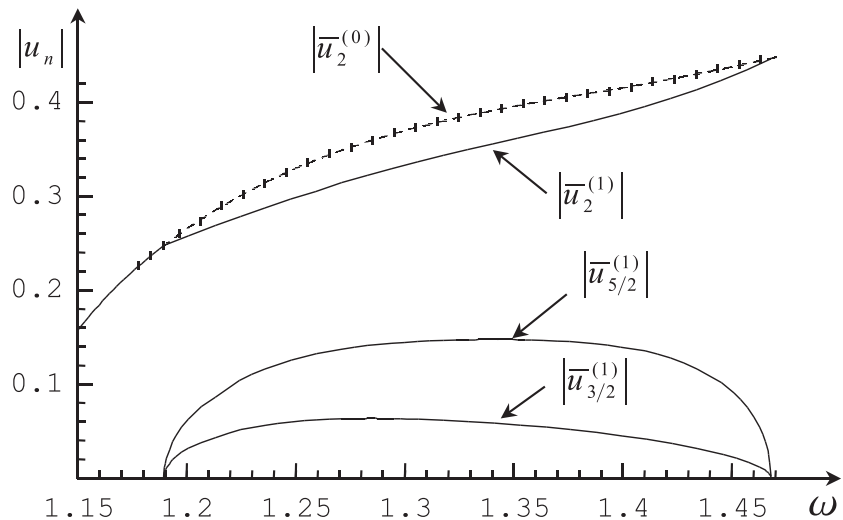

**Figure 3.** Amplitude-frequency dependences for second-order subharmonics in driven hardening Duffing's oscillator $\ddot{u} + 0.2\dot{u} + u + u^3 = 20\cos\omega t$. The harmonic balance approximation with $N = 5$ was used. ⊣⊢⊣⊢, The unstable zero level harmonic solution; ———, the stable first level subharmonic solution.

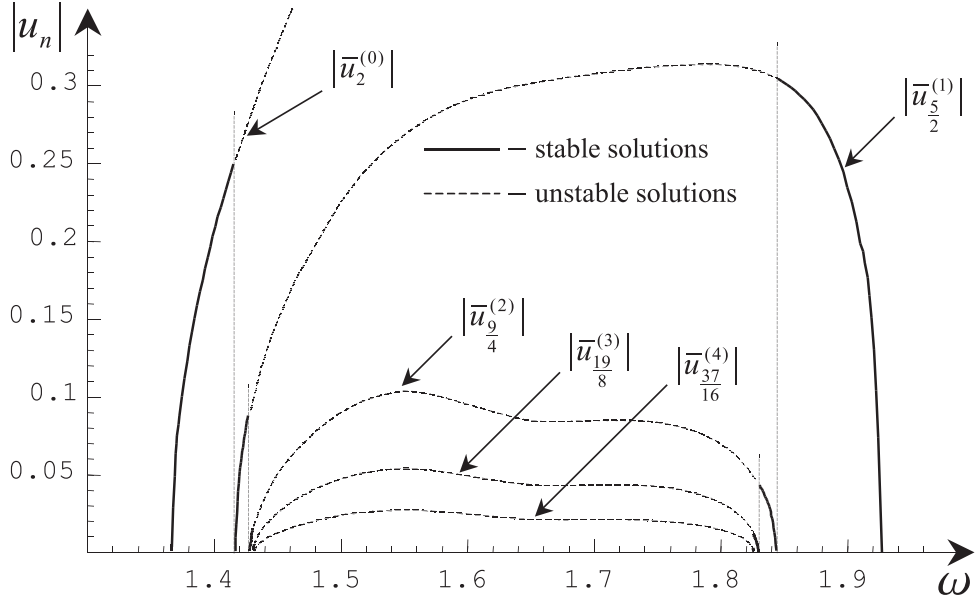

**Figure 4.** Amplitude-frequency dependences for the resonant subharmonics in the period-doubling cascade of stationary solutions to the equation $\ddot{u} + 0.2\dot{u} + u + u^3 = 40\cos\omega t$. The harmonic balance approximation with $N = 5$ was used. The unstable regions due to period-doubling are dashed.

Excitation of higher subharmonics can be observed at greater amplitudes $F$ of the external force. The cascade of period doubling $A_1 \to A_2 \to A_4 \to A_8 \to A_{16} \to \dots$ (see Figure 1) is obtained at $F = 20$. The amplitude-frequency dependences for the resonant subharmonics of stationary solutions ($N = 5$) in this cascade are shown in Figure 4. The corresponding regions of instability are the following: $\omega \in [1.4167, 1.9273]$ for the zero level solution $\bar{u}^{(0)}(t)$; $\omega \in [1.4286, 1.8450]$ for the first level solution $\bar{u}^{(1)}(t; 2)$; $\omega \in [1.4311, 1.8291]$ for the second level solution $\bar{u}^{(2)}(t; 4)$; $\omega \in [1.4316, 1.8259]$ for the third level solution $\bar{u}^{(3)}(t; 8)$; $\omega \in [1.4317, 1.8252]$ for the fourth level solution $\bar{u}^{(4)}(t; 16)$. The indicated sequence of period-doubling instabilities can be continued to higher orders, if desired. One can see that the excitation of all the subharmonics during the period-doubling bifurcations is soft. The solutions obtained are all stable only in small frequency bands in the vicinity of the bifurcation points, the width of the bands decreasing with increasing $m$.

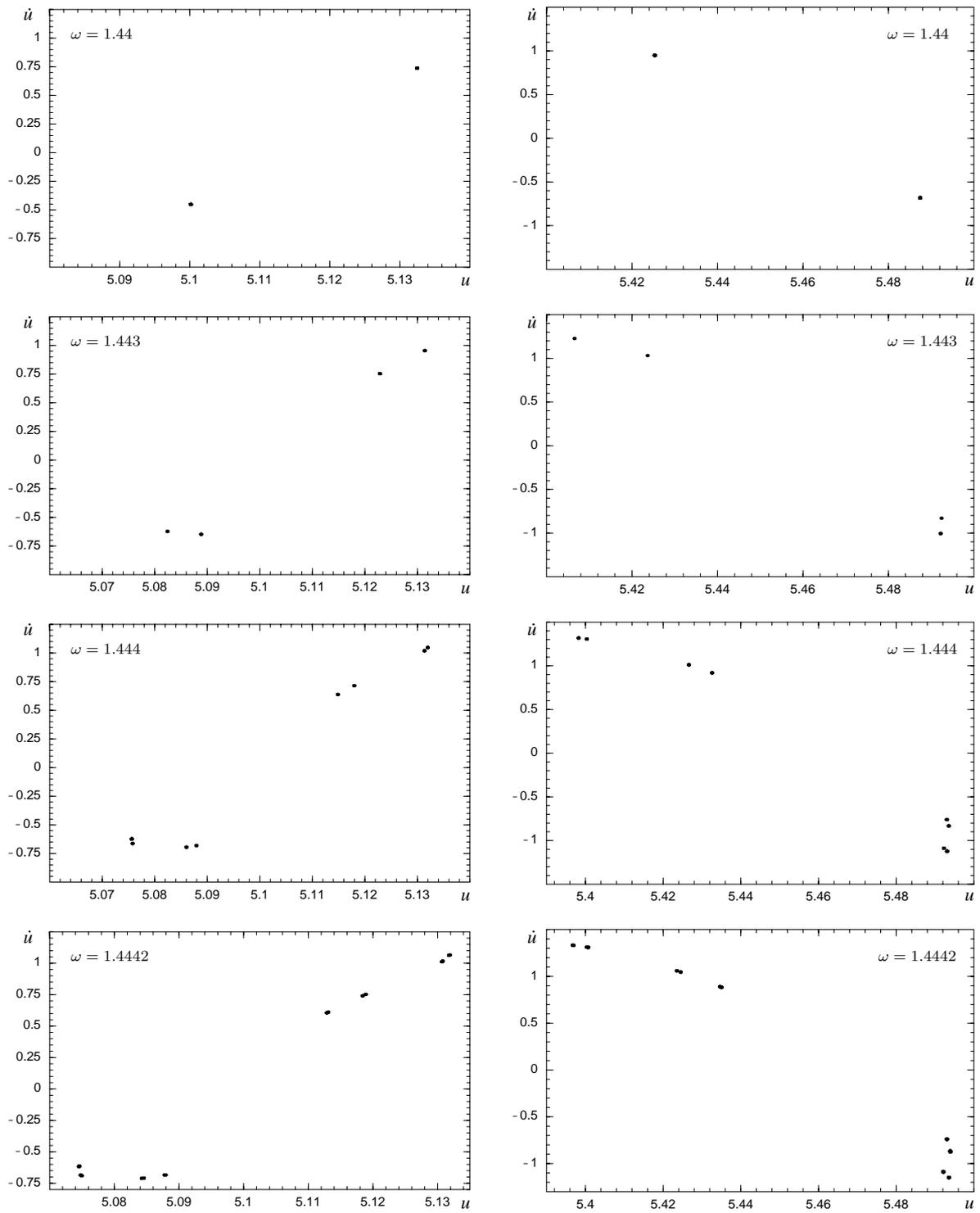

Figure 5.

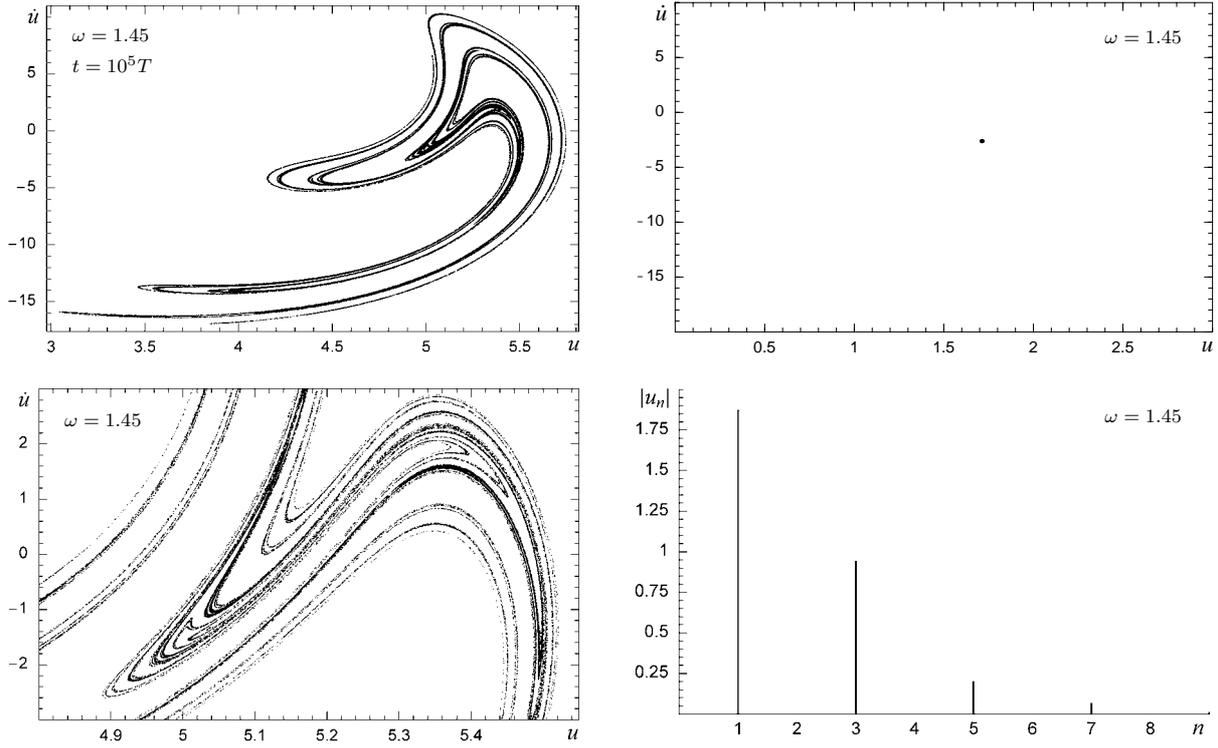

Figure 5 (continued).

**Figure 5.** The first region of the period-doubling cascade consisting of the stationary subharmonic states of order $2^n$ and leading to chaotic motion in driven hardening Duffing's oscillator $\ddot{u} + 0.2\dot{u} + u + u^3 = 40\cos\omega t$. The Poincaré maps of the following numerical subharmonic solutions are presented: $\omega = 1.44$ – $2T$-solution; $\omega = 1.443$ – $4T$-solution; $\omega = 1.444$ – $8T$-solution; $\omega = 1.4442$ – $16T$-solution; maps from the first and the second columns corresponding to two different families of asymmetric solutions shifted in phase with respect to each other. The initial conditions $u(0) = 5.075$; $\dot{u}(0) = -0.7$ and $u(0) = 5.49$; $\dot{u}(0) = -1$ were used in the fourth-order Runge-Kutta method for obtaining the solutions from the first and the second columns, respectively. The last two plots in the first column represent the chaotic attractor at $\omega = 1.45$ and its enlarged part obtained using anyone from the above initial conditions. The last two plots in the second column represent the Poincaré map and Fourier spectrum of the symmetric harmonic solution coexisting with the chaotic attractor presented, the initial condition $u(0) = 1.57$; $\dot{u}(0) = -2$ having been used.

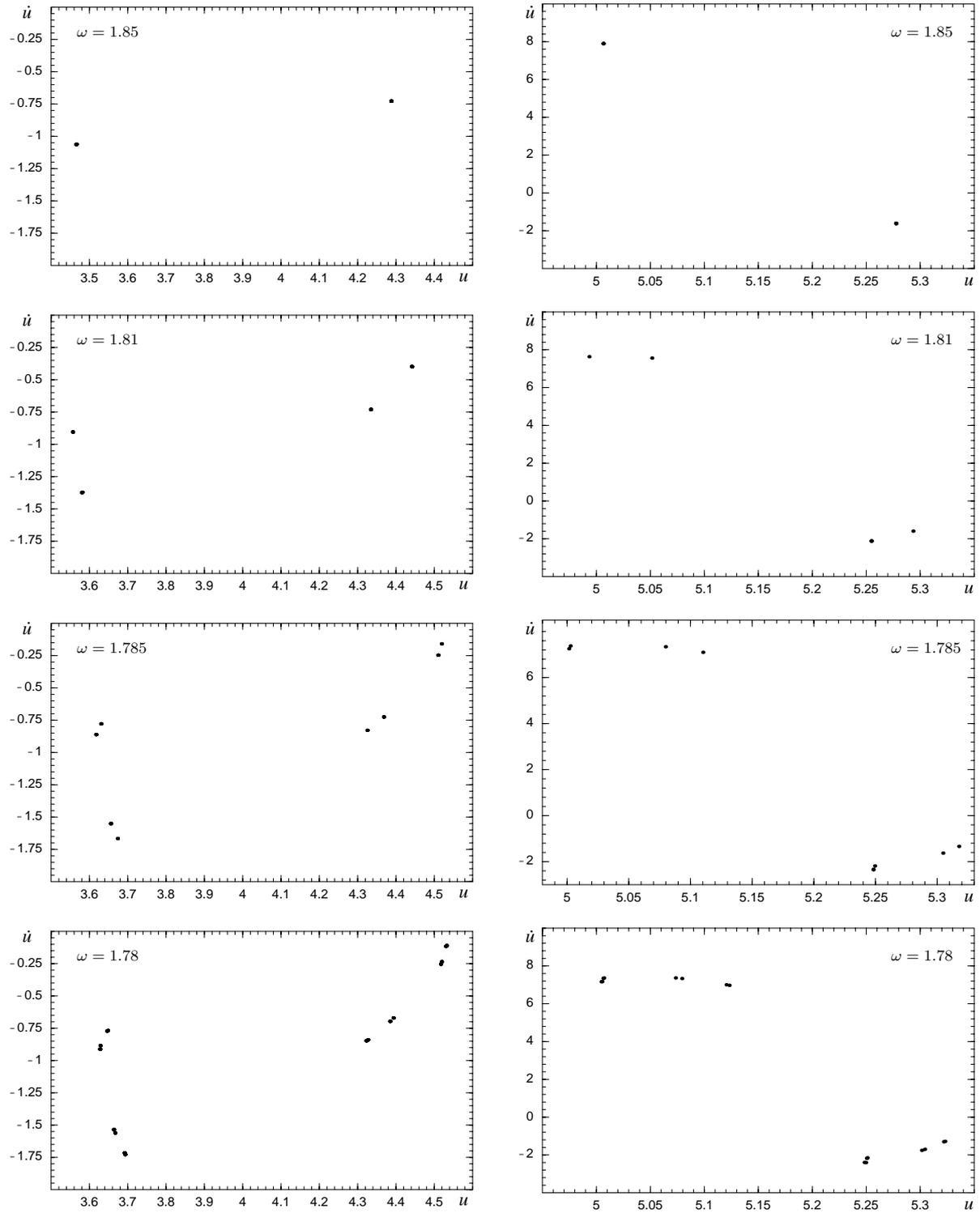

Figure 6.

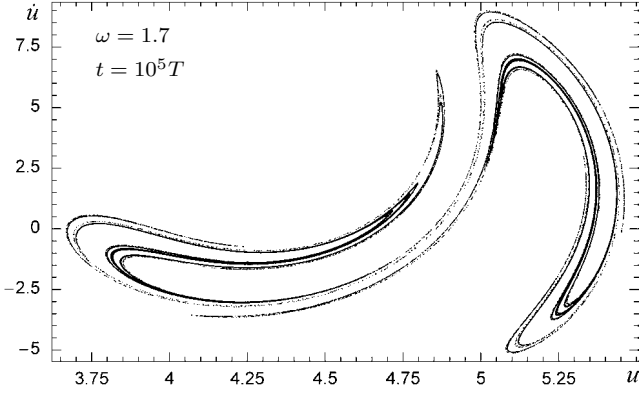

Figure 6 (continued).

**Figure 6.** The second region of the period-doubling cascade consisting of the stationary subharmonic states of order $2^n$ and leading to chaotic motion in driven hardening Duffing's oscillator $\ddot{u} + 0.2\dot{u} + u + u^3 = 40\cos\omega t$. The Poincaré maps of the following numerical stationary subharmonic solutions are presented: $\omega = 1.85 - 2T$-solution; $\omega = 1.81 - 4T$-solution; $\omega = 1.785 - 8T$-solution; $\omega = 1.78 - 16T$-solution; maps from the first and the second columns corresponding to two different families of asymmetric solutions shifted in phase with respect to each other. The initial conditions $u(0) = 3.6$; $\dot{u}(0) = -0.9$ and $u(0) = 5.25$; $\dot{u}(0) = -2$ were used in the fourth-order Runge-Kutta method for obtaining the solutions from the first and the second columns, respectively. The last plot represents the chaotic attractor at $\omega = 1.7$ obtained using anyone from the above initial conditions.

The period-doubling cascades found were confirmed to exist by solving the differential equation (3.1) ($F = 20$) numerically using the fourth-order Runge-Kutta method, the initial conditions having been all supplied by the harmonic balance approximations (3.4) at $N = 5$. The time step was chosen to achieve precision up to $10^{-10}$. The Poincaré maps of obtained stable stationary numerical solutions are shown in Figure 5 for frequencies increasing from $\omega = 1.44$ and in Figure 6 for frequencies decreasing from $\omega = 1.85$. Slight disagreement between the frequency bands for stable solutions shown in Figure 4 (obtained by the harmonic balance approximation) and the frequencies obtained from numerical simulation are only due to not a sufficient number of modes ( $N = 5$ ) used in Fourier series (3.4).

Due to symmetry (3.2) there are always two families of asymmetric solutions: $u(t; m \mid k)$ and $\hat{u}(t; m \mid k)$ (formulae (2.17) and (3.3), respectively). Because of this the even-order subharmonic solutions (they are always asymmetric) are all presented by two different Poincaré maps in Figures 5 and 6 (the Poincaré maps for the stationary solutions from the same family are the same). One can see that the second-order subharmonic solutions exist at $\omega = 1.44$ (Figure 5) and $\omega = 1.85$ (Figure 6). With increasing/decreasing frequency (Figure 5 / Figure 6) these solutions become unstable and fourth-order subharmonic solutions with doubled period exist at $\omega = 1.443$ and $\omega = 1.81$, each point in the initial Poincaré maps being splitted. In this way, subharmonic solutions of yet higher orders are formed and shown in Figures 5 and 6 at $\omega = 1.444$ and $\omega = 1.785$ (order 8); $\omega = 1.4442$ and $\omega = 1.78$ (order 16). After increasing/decreasing frequency further strange attractors were obtained at $\omega = 1.45 / \omega = 1.7$. At $\omega = 1.45$, the strange attractor coexists with the symmetric harmonic solution, whose Poincaré map and spectrum are also shown in Figure 5. It is seen from Figure 8 in Section 4, where the amplitude-frequency dependence of this symmetric solution is presented, that the third superharmonic response has not yet lost its stability (the infinite tangent has not been achieved). Therefore, originating chaotic motion is not connected with losing the stability by the third superharmonic response, as it was proposed in [24] and [15], but is due to the infinite period-doubling cascade.

## 4. THE HARD EXCITATION OF ODD-ORDER SUBHARMONIC OSCILLATIONS IN DRIVEN HARDENING DUFFING'S EQUATION

As it was noted in Section 2.3, hard excitation cannot be predicted in the framework of linear stability analysis. Therefore, calculating stationary states emerging during hard excitation is a more complicated problem. The only hint is to look for hard excitation in the same region, where soft excitation occurs. In the previous Section, only the even-order subharmonic solutions of driven hardening Duffing's oscillator (3.1) were calculated using the stability analysis. Therefore, the excitation of odd-order subharmonic oscillations is expected to be hard.

Let $\bar{u}^{(0)}(t)$ be a zero level stationary solution, e.g., the asymmetric harmonic stationary solution containing both the odd and even harmonics. Then the first level stationary solution $\bar{u}^{(1)}(t; 2m+1)$ containing the subharmonics of order $2m+1$ (being a prime number) has the following form according to (2.11) :

$$\bar{u}^{(1)}(t; 2m+1) = \bar{u}^{(0)}(t) + \bar{v}^{(1)}(t; 2m+1) = \sum_{n=-N}^{N} \left( \bar{u}_n^{(0)} + \bar{v}_n^{(1)} \right) e^{in\omega t} +$$
$$+ \sum_{n=-N}^{N-1} \sum_{n_1=1}^{2m} \bar{v}_{n+\frac{n_1}{2m+1}}^{(1)} e^{i(n+\frac{n_1}{2m+1})\omega t} = \sum_{n=-(2m+1)N}^{(2m+1)N} \bar{u}_{\frac{n}{2m+1}}^{(1)} e^{i\frac{n}{2m+1}\omega t} . \tag{4.1}$$

The harmonics $\bar{v}_{\frac{n}{m}}^{(1)}$ are found from system (2.12) at $l_1 = 1$ and $l_0 = 0$.

Indeed, the examples of such first level stationary solutions were found in equation (3.1) at $F = 20$. The amplitude-frequency dependence of the resonant subharmonic $\bar{u}_{\frac{7}{3}}^{(1)}$ of the solution $\bar{u}^{(1)}(t; 3)$ ( $N = 5$ ) is shown in Figure 7. One can see that the excitation is really hard since the amplitude of the subharmonic is nonzero at the boundaries of the interval, where it is excited. As it is seen, the obtained solution is double-valued. Its stability analysis reveals that the lower branch is metastable. And the upper branch is unstable with respect to doubling the period in the frequency band $\omega \in [1.4070, 1.9214]$. In this region, the second level stationary solution $\bar{u}^{(2)}(t; 6)$ is softly excited and is presented by the greatest amplitude $|\bar{u}_{\frac{13}{6}}^{(2)}|$ in Figure

7. This second level solution undergoes subsequent period-doubling in the frequency band $\omega \in [1.4074, 1.9202]$. The obtained cascade corresponds to the sequence $A_1 \to A_3 \to A_6$ in Figure 1.

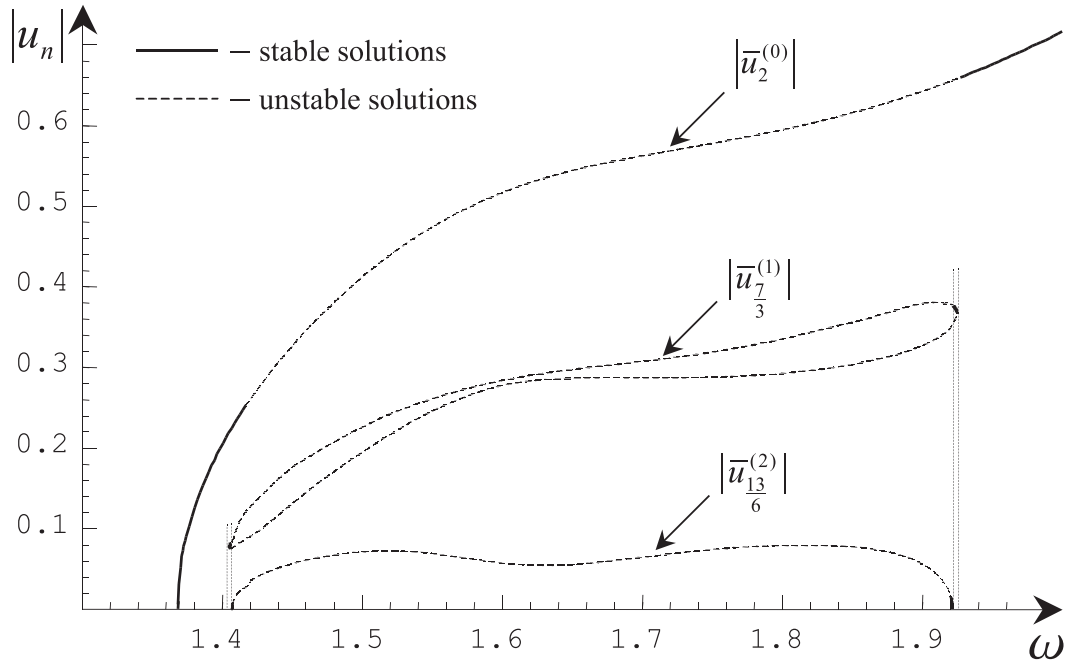

**Figure 7.** Amplitude-frequency dependences for the resonant subharmonics of both the asymmetric third-order subharmonic solution that is hardly excited and the sixth-order subharmonic solution that is softly excited due to subsequent period-doubling in driven hardening Duffing's oscillator $\ddot{u} + 0.2\dot{u} + u + u^3 = 40\cos\omega t$. The harmonic balance approximation with $N = 5$ was used. The unstable regions are dashed.

In this Section, however, main attention is devoted to symmetric subharmonic solutions of odd orders containing only the odd subharmonics ($\overline{u}^{(1)}_{\frac{2n}{2m+1}} = 0$). They can be obtained similarly to (4.1) as the deviations to the zero level solution $\tilde{u}^{(0)}(t)$ containing only the odd harmonics. The tilde means that a solution is stationary and the even harmonics are all zero. The amplitude-frequency dependences for the resonant subharmonics of such first level stationary solutions $\tilde{u}^{(1)}(t;3)$ and $\tilde{u}^{(1)}(t;5)$ ($N = 5$) to equation (3.1) with $F = 20$ are shown in Figure 8 (the solution $\tilde{u}^{(1)}(t;3)$ was also obtained in [22] at $F = 25$). The obtained solutions are double-valued as in Figure 7. This seems to be the universal property of the solutions that are hardly excited. The lower branches are metastable. And the upper branches are unstable in certain regions with respect to excitation of the even subharmonics of the same order (symmetry-breaking), namely: the solution $\widetilde{u}^{(1)}(t;3)$ is unstable at $\omega \in [1.3619, 1.5325]$; the solution $\widetilde{u}^{(1)}(t;5)$ is unstable at $\omega \in [1.4046, 1.8850]$. These bands, however, are not precise enough and will slightly change after taking into account a greater number of harmonics.

The solution $\widetilde{u}^{(1)}(t;3)$ has one more peculiarity. At the point $C$ in Figure 8, the upper stable and the lower metastable branches intersect and exchange their stability due to a transcritical bifurcation. The upper stable solution becomes metastable (as during a saddle-node bifurcation) but without an infinite tangent and, vice versa, the lower metastable solution becomes stable.

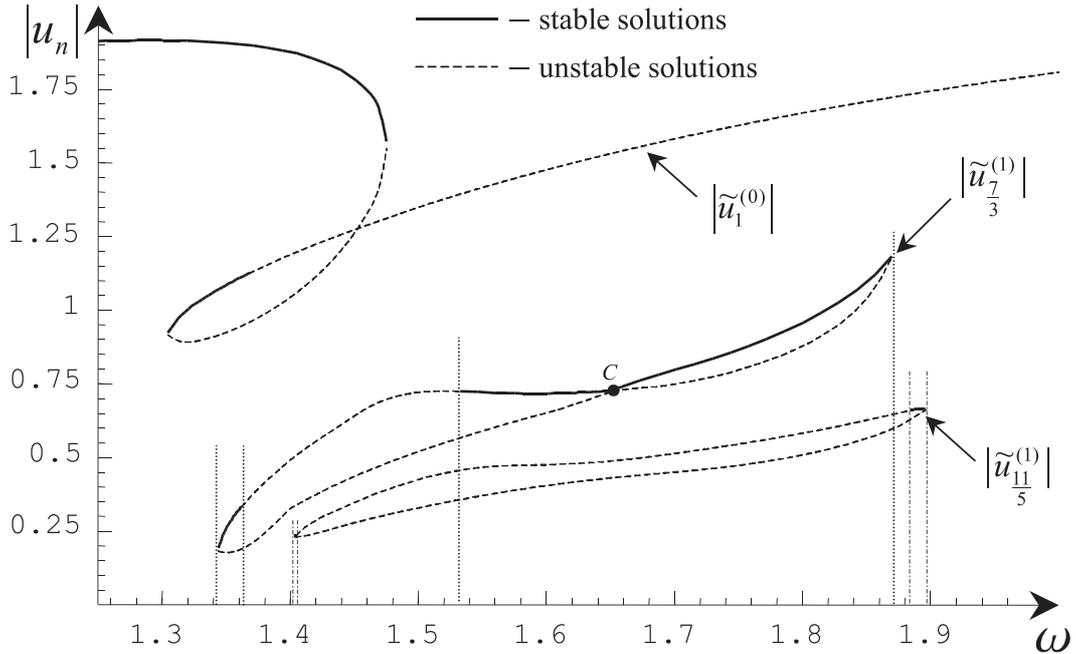

**Figure 8.** Amplitude-frequency dependences for the resonant subharmonics of the symmetric third-order and fifth-order subharmonic solutions that are hardly excited in driven hardening Duffing's oscillator $\ddot{u} + 0.2\dot{u} + u + u^3 = 40\cos\omega\,t$. The harmonic balance approximation with $N = 5$ was used. The unstable regions are dashed.

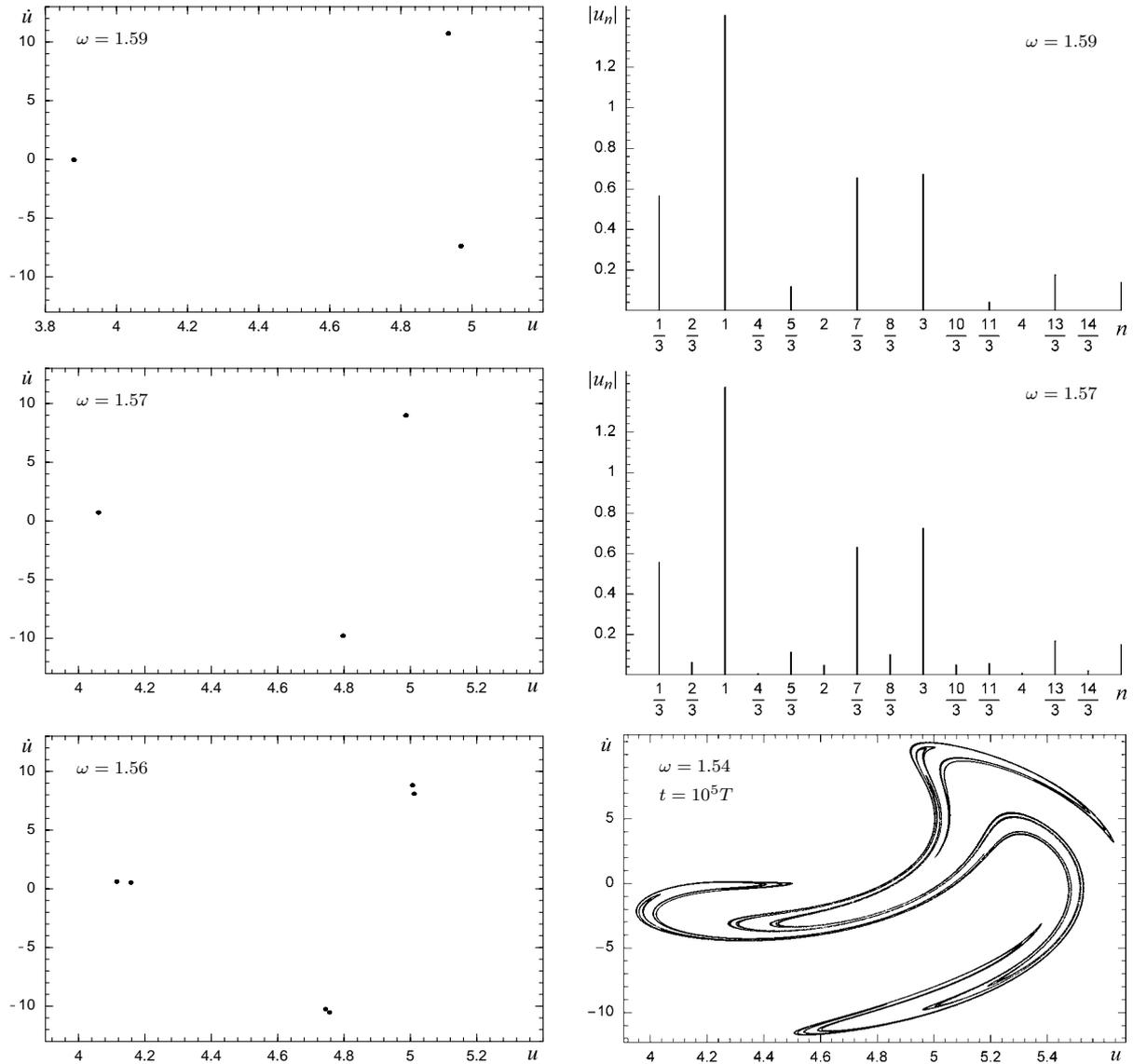

Figure 9.

**Figure 9.** The period-doubling cascade starting from hardly excited symmetric third-order subharmonic solution and leading to chaotic motion in driven hardening Duffing's oscillator $\ddot{u} + 0.2\dot{u} + u + u^3 = 40\cos\omega t$. The Poincaré maps of the following stationary subharmonic solutions are presented: $\omega = 1.59$ – symmetric $3T$-solution; $\omega = 1.57$ – asymmetric $3T$-solution; $\omega = 1.56$ – $6T$-solution; the Fourier spectra of both $3T$-solutions being also presented. The last plot represents the chaotic attractor at $\omega = 1.54$. The initial condition $u(0) = 4.75$; $\dot{u}(0) = -10$ was used in the fourth-order Runge-Kutta method for obtaining all the solutions.

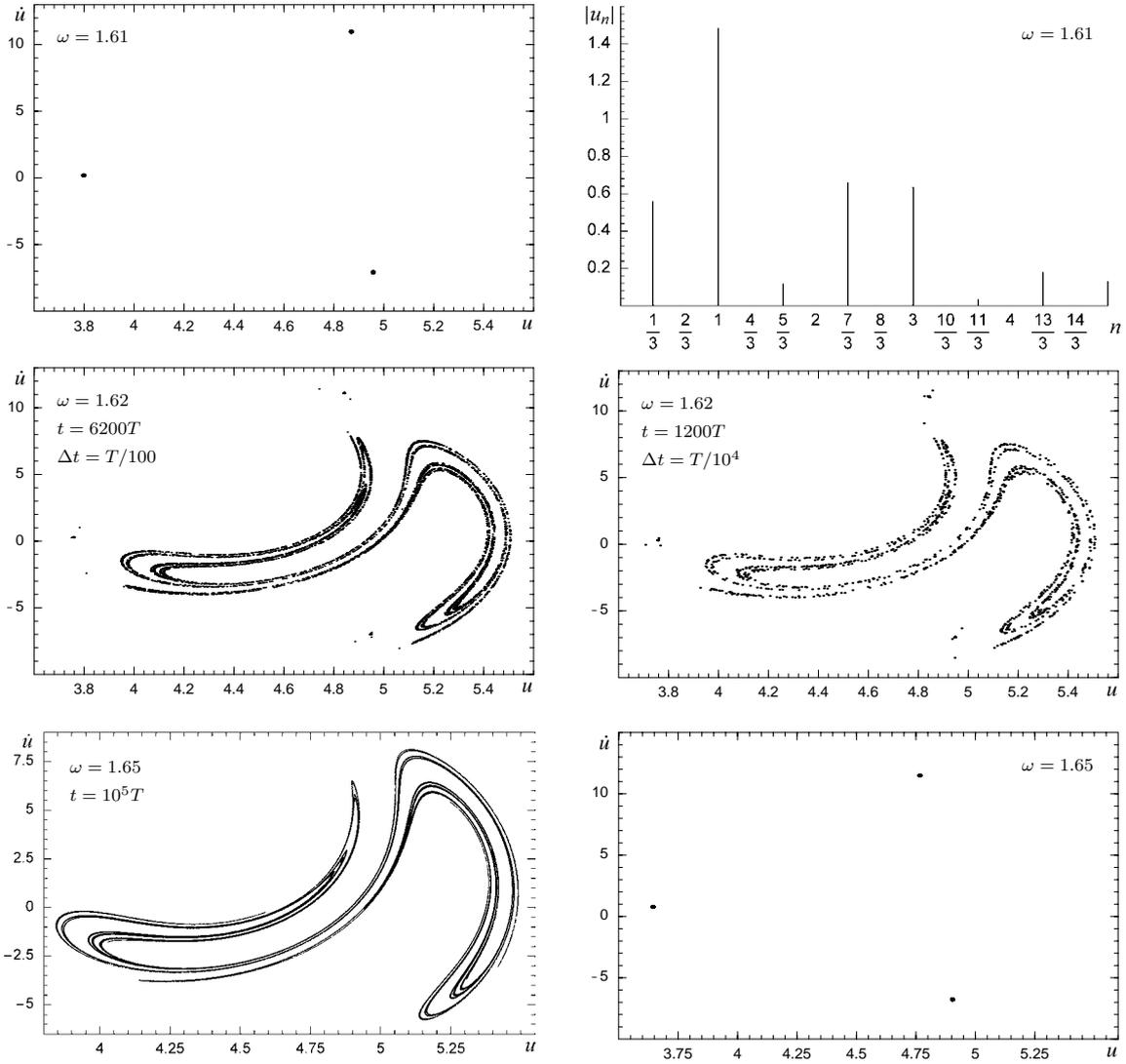

Figure 10.

**Figure 10.** Abrupt transition to chaotic motion via transient chaos without a period-doubling cascade in driven hardening Duffing's oscillator $\ddot{u} + 0.2\dot{u} + u + u^3 = 40\cos\omega\,t$. The first row: the Poincaré map and Fourier spectrum of the symmetric third-order subharmonic solution being the only steady solution at $\omega = 1.61$. The second row ($\omega = 1.62$): transient chaos that settles down onto the symmetric third-order subharmonic solution after elapsing the indicated period of time, the Poincaré maps being presented for two different time steps $\Delta t$ in the fourth-order Runge-Kutta method used. The third row: the full-time chaotic attractor and the symmetric third-order subharmonic solution, which coexist at $\omega = 1.65$. The initial conditions $u(0) = 4.95$; $\dot{u}(0) = -7$ and $u(0) = 5.25$; $\dot{u}(0) = -2$ were used for obtaining the subharmonic solutions and chaotic attractors, respectively.

On the basis of the data obtained using the harmonic balance approximation, numerical calculations were carried out using the fourth-order Runge-Kutta method similarly to Section 3.2. The period-doubling cascade starting from the symmetric solution $\tilde{u}^{(1)}(t; 3)$ is presented in Figure 9 by using the Poincaré maps and Fourier spectra. At $\omega = 1.59$, $\tilde{u}^{(1)}(t; 3)$ is the only stable stationary solution to equation (3.1) at $F = 20$. After decreasing frequency a symmetry-breaking bifurcation occurs, with even harmonics being excited, as it was predicted by the stability analysis of the harmonic balance approximation. Such an asymmetric third-order subharmonic solution is presented in Figure 9 at $\omega = 1.57$. After decreasing frequency further the sixth-order subharmonic solution is found at $\omega = 1.56$. Note that only one family of asymmetric solutions is presented, with the second one having been omitted. The infinite sequence of period-doubling bifurcations leads to chaotic motion found at $\omega = 1.54$.

Thus, chaotic motion originating after the infinite period-doubling cascade at $\omega \approx 1.45$ (see Figure 5) is continued when increasing frequency up to $\omega \approx 1.54$ and then disappears, with the reversed period-doubling cascade leading to the hardly excited symmetric solution $\tilde{u}^{(1)}(t; 3)$ (see Figure 9 from the bottom to the top). This is the only steady solution starting from $\omega \approx 1.59$ up to $\omega \approx 1.61$ (see the first row of the plots in Figures 9 and 10). Such a solution is usually called a periodic window inside the region of chaotic motion [9]. Thus, appearance of periodic windows is due to hardly excited stable subharmonic solutions.

At $\omega \approx 1.62$, chaotic motion returns to the system in the form of transient chaos, as one can see from Figure 10. Transient chaotic motion settles down onto the stable solution $\tilde{u}^{(1)}(t; 3)$ after elapsing some period of time. The region of transient chaos abruptly finishes at $\omega \approx 1.62011$ and full-time chaotic motion coexisting with the solution $\tilde{u}^{(1)}(t; 3)$ is observed for greater frequencies, as is shown in Figure 10 for $\omega = 1.65$, up to the threshold ( $\omega \approx 1.75$ ) of the period-doubling cascade presented in Figure 6 (look from the bottom to the top). The route to/from chaotic motion via transient chaos occurs without a period-doubling cascade and is known as a boundary crisis [9]. The threshold for chaotic motion in this case to all appearance corresponds to the transcritical point $C$ in Figure 8, where the stable and unstable branches of the solution $\widetilde{u}^{(1)}(t;3)$ intersect and exchange their stability.

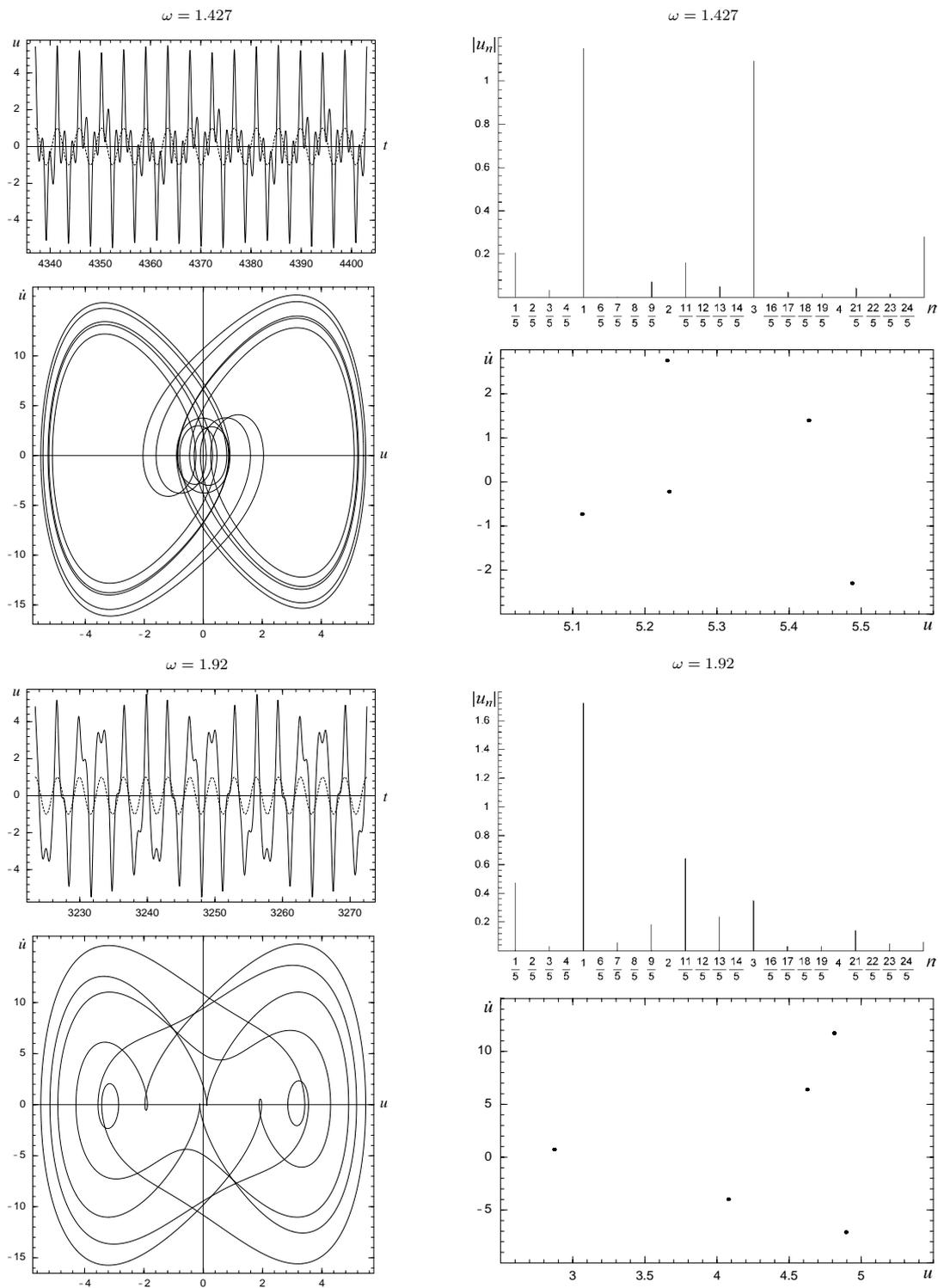

Figure 11.

**Figure 11.** The time evolution, Fourier spectra, phase portraits, and Poincaré maps of the symmetric fifth-order subharmonic solution obtained using the fourth-order Runge-Kutta method at $\omega = 1.427$ ( $u(0) = 5.427$ ; $\dot{u}(0) = 1.397$ ) and $\omega = 1.92$ ( $u(0) = 4.81$ ; $\dot{u}(0) = 11.73$ ).

Finally, the stable symmetric solution $\widetilde{u}^{(1)}(t;5)$ is numerically found at $\omega = 1.427$ and $\omega = 1.92$ that is also in good agreement with Figure 8. The corresponding time evolution, Fourier spectra, phase portraits, and Poincaré maps are shown in Figure 11. Time evolution and phase portraits of the solution $\widetilde{u}^{(1)}(t;5)$ from the different sides ($\omega = 1.427$ and $\omega = 1.92$) of its amplitude-frequency dependence (see Figure 8) are quite different. The subharmonic solution at $\omega = 1.427$ is only a modulated harmonic solution, whereas the solution at $\omega = 1.92$ yields new oscillatory and phase portraits.

## 5. CONCLUSIONS

In this paper, we proposed a consecutive scheme for studying harmonic and subharmonic driven oscillations described by second-order differential equations with arbitrary polynomial non-linearity. The technique is based on the harmonic balance method and Floquet theorem adapted for calculating stationary subharmonic solutions of arbitrary order for arbitrary truncation numbers and investigating the stability of such general solutions. To classify cascades of subharmonic solutions with increasing orders a general diagram is proposed, where the possible routes for obtaining subharmonic solutions of higher orders are all taken into account.

The effectiveness of the proposed technique was demonstrated on driven hardening Duffing's oscillator (3.1), with a variety of stable and unstable subharmonic stationary solutions having been obtained. Using initial conditions supplied by the harmonic balance approximation the results were numerically verified by means of the fourth-order Runge-Kutta method.

## ACKNOWLEDGEMENTS

This research was supported by INTAS Grant № 99–1637. The research of I. Gandzha has also been supported by INTAS YSF 2001/2–114.

Extraneous solutions predicted by the harmonic balance method.

## APPENDIX A. Fourier harmonics of the product of two periodic functions

Let the periodic functions $u(t)$, $v(t)$, and their product be approximated by the following Fourier series with taking into account the subharmonics of order $m$:

$$u(t) = \sum_{n=-mN_u}^{mN_u} u_{\frac{n}{m}} e^{i\frac{n}{m}\omega t}; \quad v(t) = \sum_{n=-mN_v}^{mN_v} v_{\frac{n}{m}} e^{i\frac{n}{m}\omega t}; \quad u(t)\cdot v(t) = \sum_{n=-m(N_u+N_v)}^{m(N_u+N_v)} (uv)_{\frac{n}{m}} e^{i\frac{n}{m}\omega t}. \quad \text{(A1)}$$

Then the harmonics $(uv)_{\frac{n}{m}}$ are expressed in terms of the harmonics $u_{\frac{n}{m}}$ and $v_{\frac{n}{m}}$ by the following convolution

$$(uv)_{\frac{n}{m}} = \sum_{n_1=\max(-mN_u,\,n-mN_v)}^{\min(mN_u,\,n+mN_v)} u_{\frac{n_1}{m}} v_{\frac{n-n_1}{m}}, \quad |n| = \overline{0,\, N_u+N_v}. \quad \text{(A2)}$$

The following explicit form for harmonics of the periodic function $v^p$, $p$ being an arbitrary integer, can be obtained using (A2):

$$(v^p)_{\frac{n}{m}} = \sum_{n_1=N_1^-}^{N_1^+} v_{\frac{n-n_1}{m}} \cdots \sum_{n_i=N_i^-}^{N_i^+} v_{\frac{n_{i-1}-n_i}{m}} \cdots \sum_{n_{p-1}=N_{p-1}^-}^{N_{p-1}^+} v_{\frac{n_{p-2}-n_{p-1}}{m}} v_{\frac{n_{p-1}}{m}}, \quad |n| = \overline{0,\, pN}; \quad \text{(A3)}$$

$$N_i^- = \max(n_{i-1}-mN,\,-(p-i)mN), \quad N_i^+ = \min(n_{i-1}+mN,\,(p-i)mN).$$

## APPENDIX B. An eigenvalue problem for characteristic frequencies

The system of second-order differential equations (2.9) can be reduced to a system of first-order equations by introducing the variables $w_{\frac{n}{m}}^{(l_1)}(t) \equiv \dot{v}_{\frac{n}{m}}^{(l_1)}(t)$. Linearization of this system of equations with taking into account the Floquet theorem $v_{\frac{n}{m}}^{(l_1)}(t) \equiv v_{\frac{n}{m}}^{(l_1)} \exp(i\Omega t)$, $w_{\frac{n}{m}}^{(l_1)}(t) = w_{\frac{n}{m}}^{(l_1)} \exp(i\Omega t)$ leads to the following eigenvalue problem for $\lambda = -i\Omega$:

$$i\Omega v_{\frac{n}{m}}^{(l_1)} - w_{\frac{n}{m}}^{(l_1)} = 0, \quad |n| = \overline{0,\, mN};$$

$$\sum_{n_1=N_n^-(1;\,m)}^{N_n^+(1;\,m)} \left( \left( f_{\frac{n_1}{m}} + \alpha_1 \right) \delta_{n,\,n_1} + \beta_{1,\,\frac{n-n_1}{m}}^{(l_0)} \right) v_{\frac{n_1}{m}}^{(l_1)} + (i\Omega + 2i\frac{n}{m}\omega + g) w_{\frac{n}{m}}^{(l_1)} = 0, \quad |n| = \overline{0,\, mN}. \quad \text{(B1)}$$